\documentclass[preprint,showpacs,preprintnumbers,amsmath,amssymb]{revtex4}
\usepackage{graphicx}
\usepackage{dcolumn}
\usepackage{bm}
\begin{document}

\title{Pulverization of the flux line lattice, the phase coexistence and the spinodal temperature of the order-disorder transition in a weakly pinned crystal of Yb$_3$Rh$_4$Sn$_{13}$}
\author{S Sarkar$^1$, C V Tomy$^{2,\ast}$, A  D  Thakur$^1$, G Balakrishnan$^3$, D McK Paul$^3$, S  Ramakrishnan$^1$ and A  K  Grover$^{1,\ast\ast}$}
\affiliation{$^1$ DCMP\&MS, Tata Institute of Fundamental Research, Mumbai
400 005, India \\
$^2$ Department of Physics, Indian Institute of Technology Bombay, Mumbai
400 076, India \\
$^3$ Department of Physics, University of Warwick, Coventry CV4 7AL, UK\\
}

\date{today}

\begin{abstract}
We have studied metastability effects pertaining to the peak effect (PE) in critical current density ($J_c$) via isofield scans in ac susceptibility measurements in a weakly pinned single crystal of Yb$_3$Rh$_4$Sn$_{13}$ ($T_c(0) \approx 7.6 $~K). The order-disorder transition in this specimen proceeds in a multi-step manner. The phase coexistence regime between the onset temperature of the PE and the spinodal temperature (where metastability effects cease) seems to comprise two parts, where ordered and disordered regions dominate the bulk behavior, respectively. The PE line in the vortex phase diagram is argued to terminate at the low field end at a critical point in the elastic (Bragg) glass phase.
\end{abstract}



\maketitle

\section{Introduction}

The second last element of the rare earth series, Yb, easily lends itself to a mixed valent character (Yb$^{3+}$ to Yb$^{2+}$) in intermetallic compounds and hence the Yb based compounds often exhibit anomalous magnetic behavior, analogous to those observed in Ce and U based compounds. The ternary intermetallic compound, Yb$_3$Rh$_4$Sn$_{13}$ in the primitive cubic phase I ($Pm3n$) structure, exhibits superconductivity in the temperature range 7 to 8~K \cite{1}. Electrical, magnetic and specific heat studies pertaining to the peak effect phenomenon in critical current density in weakly pinned single crystals of Yb$_3$Rh$_4$Sn$_{13}$ ($T_c(0) \approx$ 7.6~K) were undertaken by Sato {\it et al} \cite{2} in the context of assertions made relating to the realization of the Fulde-Ferrel-Larkin-Ovchinnikov (FFLO) state in certain Ce and U compounds, e.g., UPd$_2$Al$_3$, \cite{3,4}, UPt$_3$ \cite{5}, CeCo$_2$ \cite{6}, CeRu$_2$ \cite{7,8}, etc. Tomy {\it et al} \cite{9} also studied the PE phenomenon in their single crystals of Yb$_3$Rh$_4$Sn$_{13}$ ($T_c(0) \approx$ 7.6~K) via isothermal magnetization hysteresis and magneto-resistance measurements and surmised that the observed behavior was akin to that in CeRu$_2$ \cite{7,10,11,12}. Later, Tomy {\it et al} \cite{13} investigated PE phenomenon in a single crystal of Ca$_3$Rh$_4$Sn$_{13}$ ($T_c(0) \approx$ 8.2~K), a compound isostructural to Yb$_3$Rh$_4$Sn$_{13}$, and found a behavior similar to that in the latter. However, the Ca compound with divalent non-magnetic Calcium ions cannot be construed to be a candidate for the FFLO state. A later exploration of the PE phenomenon in single crystals of 2$H$-NbSe$_2$ and CeRu$_2$ by Banerjee {\it et al} \cite{14} revealed the notion of fracturing of the ordered vortex lattice across the PE region in isofield ac susceptibility scans. This study disfavored the necessity of invoking the FFLO state for the PE phenomenon in CeRu$_2$. Sarkar {\it et al} \cite{15,16} reported the presence of two-step amorphization in isofield ac susceptibility measurements in a weakly pinned single crystal of Ca$_3$Rh$_4$Sn$_{13}$. Subsequently, Sarkar {\it et al} \cite{17} also found a multi-step amorphization of the ordered vortex solid across the PE region in a single crystal of Yb$_3$Rh$_4$Sn$_{13}$. A notable difference in the sequence of fracturing steps across the PE region in the case of Ca and Yb based crystals studied by Sarkar {\it et al} \cite{15,17} was that in the Yb sample, the fracturing path is partially retraceable (cf. Fig.~2 in Ref. \cite{17}) whereas in the Ca sample, once the fracturing gets triggered in the ordered lattice, the attempts to retrace the path lead to a progress towards a supercooled disordered state. It had been stated that in the Yb case \cite{17}, some of the dislocations that invade the sample above the onset temperature of the PE can be squeezed out in the process of cooling down, while the rest do not get driven out without an additional an external stimulus. A reassessment of this observation in the light of the subsequent revelations \cite{18,19,20} about the details of the happenings across the PE region in weakly pinned crystals of $2H$-NbSe$_2$ implies that the temperature interval of the anomalous variation in $J_c$ comprises the coexistence of ordered and disordered pockets. Such a coexistence region can be subdivided into two parts, in which the ordered and disordered regions dominate, respectively.

We report here the details of our ac susceptibility measurements performed on a single crystal of Yb$_3$Rh$_4$Sn$_{13}$, which substantiate the above stated scenario. Application of an external stimulus (even for a short duration) can in fact wash away the initial few steps of the pulverization process and provide newer paths to amorphization of the ordered vortex state across the latter part of the PE region. Retracing the temperature from various points in the coexisting order-disorder phase highlights the evidence supporting the bifurcation of the coexistence region, akin to that in the case of $2H$-NbSe$_2$ system \cite{19,20}.      

\section{Experimental details}

The isofield ac susceptibility measurements were performed on a single crystal sample (mass = 140~mg) of Yb$_3$Rh$_4$Sn$_{13}$,  grown  by the tin flux method by Tomy et al. \cite{9}. This sample has a $T_c(0)$ of 7.6~K ($\Delta T_c(0) \approx$ 50~mK). The ac susceptibility measurements in superposed dc magnetic fields have been carried out using a well-shielded
home built ac susceptometer \cite{21}. The ac and dc fields were co-axial and the sample was placed in such a way that one of its principal axis (cube edge) was aligned parallel with
the field (i.e., $H$~//~[001]). Most of the ac susceptibility data were usually made at a frequency of 211~Hz and with an ac amplitude (r.m.s.) of 0.65~Oe.

\section{Results and Discussion}

\subsection{Surfacing of the Peak Effect in the isofield ac susceptibility measurements}

The plots in Fig.~1 focus attention onto the surfacing of the PE phenomenon in the isofield in-phase ac susceptibility ($\chi^{\prime}(T)$) data as the applied field is progressively enhanced from 4~kOe to 6~kOe. The main panel shows the $\chi^{\prime}(T)$ plot in a field of 6~kOe, where the PE manifests itself as an easily recognizable dip in the ac shielding response. The $\chi^{\prime}(T)$ is known to relate to $J_c$ via the equation \cite{22}, $\chi^{\prime}(T) \sim -1 + \alpha h_{ac} / J_c$, where, $\alpha$ is sample size and geometry dependent factor and $h_{ac}$ is the amplitude of the ac drive. The onset of a dip in $\chi^{\prime}(T)$ implies a PE like feature in $J_c$. Note that the onset of the PE fingerprints as a  (sharp) discontinuous change in $\chi^{\prime}(T)$ response at $T \approx T_{pl}$. It is considered that the vortex lattice is elastically pinned prior to $T_{pl}$, the permeation of dislocations  at $T \approx T_{pl}$ introduces plastic deformations which continue in a multi-step manner up to the notional peak temperature ($T_p$) in $\chi^{\prime}(T)$ response. Above $T_p$, the sharp collapse in the pinning precipitates. The PE phenomenon thus presents itself as a change in the form from an elastic to plastic vortex solid as a consequence of the interplay between the collapse of the elasticity and the pinning as $T \rightarrow T_c(H)$ \cite{23}. The inset panels (a) and (b) in Fig.~1 display $\chi^{\prime}(T)$ plots in fields of 4~kOe and 5~kOe, respectively. While the plot at 5~kOe in panel (b) clearly imbibes the notion of the PE anomaly, the plot at 4~kOe in panel (a) displays only a shallow modulation across the temperature marked as $T_p$ in this panel. This shallow modulation can be construed as an anomalous variation in $J_c(T)$ across $T_p$ at $H = 4$~kOe. Note that the $\chi^{\prime}(T)$ responses at temperatures before and after this anomaly would appear to continuously connect to each other, if the anomaly in $J_c$ is absent. This behavior is different from $\chi^{\prime}(T)$ response before and after the PE anomaly at 5~kOe and 6~kOe as shown in the inset panel (b) and the main panel, respectively. In the latter cases, the surfacing of the robust PE is accompanied by a sharper collapse in pinning, concurrent with the (notional) peak position of the PE. Any anomalous variation in $\chi^{\prime}(T)$ response could not be identified in our measurements in  Yb$_3$Rh$_4$Sn$_{13}$ for applied fields $H < 4$~kOe (the data for 3~kOe is not shown here). At $H = 4$~kOe, the intervortex spacing for a triangular FLL, $a_0$ is about 800~\AA, which is well within the estimated penetration depth ($\lambda$) value of 2100~\AA \cite{2} in this compound. Usually, one would expect the FLL to be well formed when $a_0 < \lambda$. The non-observation of the PE for fields in the interval 0.5~kOe $<$ $H$ $<$ 3~kOe, therefore, desires a plausible explanation. An analogous situation had been noted earlier in the context of $\chi^{\prime}(T)$ data in a single crystal of Ca$_3$Rh$_4$Sn$_{13}$ \cite{16}, wherein the limiting field value at which the fingerprint of an anomaly in $J_c$ could be  first noticeable at was $H = 3.5$~kOe (corresponding to $a_0 \approx$ 830~\AA), whereas its $\lambda$ value was estimated to be 2270~\AA at 4.5~K.

\subsection{Path dependence in the $\chi^{\prime}(T)$ response at $H = 8$~kOe}

The inset panel in Fig.~2 shows $\chi^{\prime}(T)$ data in a field of 8~kOe in the ZFC state with  $h_{ac}$ = 0.65~Oe. The main panel of Fig.~2 shows a comparison of the $\chi^{\prime}(T)$ response in the ZFC state with those in the field cool cool-down (FCC) and the field cool warm-up modes (FCW) on an expanded scale near the notional peak position of the PE. The full $\chi^{\prime}(T)$ curve in the inset panel reveals the occurrence of several modulations above the onset temperature (designated as $T_{pl1}$ in Ref. \cite{17}) of the PE anomaly. In the main panel of Fig.~2, we have identified the different locations ($T_{pl1}$ to $T_{pl4}$) of the modulations/step changes in the $\chi^{\prime}(T)$ response in the ZFC mode. The occurrence of discontinuous change in the $\chi^{\prime}(T)$ response can be noted across the temperatures $T_{pl2}$ to $T_{pl4}$ in the FCC and FCW modes as well. Above $T_{pl4}$, the diamagnetic $\chi^{\prime}(T)$ response in all the three modes appears to rapidly decrease and the history dependence in $\chi^{\prime}(T)$ response ceases somewhat above the temperature $T_{pl4}$. Such a limiting temperature has been identified as $T^{\star}$ in Fig.~2 (see further discussion in section 3.3).

\subsection{Effect of an ac driving force ($h_{ac}$) on $\chi^{\prime}(T)$ response at $H = 8$~kOe}

Figure~3 shows a comparison of the $\chi^{\prime}(T)$ responses in a field of 8~kOe recorded at different amplitudes of the ac field ($h_{ac}$), ranging from 0.65~Oe to 2~Oe. The $h_{ac}$ performs a dual role in ac susceptibility measurements. While invoking the critical currents in the vortex state, $h_{ac}$ could also shake the vortices and drive the system towards the equilibrium state at a given ($H, T$). The plots in Fig.~3 show that prior to the onset temperature of the PE (marked by arrows as $T_{pl1}$), the larger $h_{ac}$ fields drive the vortex state towards lower $J_c$ values (i.e., more ordered states). However, after the onset temperature, as the disordering sets in, the vortex states (having differing spatial order) rapidly move towards the most disordered configuration at 8~kOe, and eventually the rapid depinning commences. It is interesting to note that the fracturing temperatures, $T_{pl3}$ and $T_{pl4}$, at which the sharp drops in $\chi^{\prime}(T)$ happen, can be noted in each of the curves of Fig.~3. A similar trend had been (cf. Fig.~1 in \cite{15}) noted in the  $\chi^{\prime}(T)$ data in a single crystal of Ca$_3$Rh$_4$Sn$_{13}$ in a field of 10~kOe, where the two fracturing temperatures were found to be independent of $h_{ac}$ and the frequency of the ac field (in the interval 21~Hz to 211~Hz). The $\chi^{\prime}(T)$ response is seen to become independent of $h_{ac}$ at a limiting value $T^{\star}$, which is greater than $T_{pl4}$. It appears superfluous to assign special significance to the peak temperature in the $\chi^{\prime}(T)$ response in the context of data in Yb$_3$Rh$_4$Sn$_{13}$ at a field of 8~kOe. The $\chi^{\prime}(T)$ curves in different thermomagnetic histories (cf. Fig.~2) would imply different peak temperatures for different modes, such that $T_p^{FCC} > T_p^{ZFC}$. Analogous trend had recently been noted from the analysis of the data in the well studied $2H$-NbSe$_2$ system \cite{19,20}. The peak temperature even in the ZFC mode does not identify the limiting temperature at which the thermomagnetic history effects in $J_c(T)$ for a given $H$ cease. Such a temperature lies above $T_p^{ZFC}$ and is probably the $T^{\star}$, as marked in Fig.~2 and Fig.~3. If we invoke the Larkin-Ovchinnikov relationship \cite{24} between $J_c$ and the volume $V_c$ of the domain in which FLL is correlated ($J_c \propto 1/ \surd V_c$), the $\chi^{\prime}(T)$ data in Fig.~2 would imply that the vortex state prepared at 8~kOe in the ZFC mode has larger correlation volume than those prepared in the FCC or FCW modes for $T < T_{pl1}$. The $\chi^{\prime}(T)$ curves in Fig.~3 further imply that the ZFC state under the continuous influence of a higher $h_{ac}$ drive has even larger correlation volume up to $T_{pl3}$. These observations raise a query, whether the state of spatial order in the initially prepared ZFC state can be further improved by an exposure to a driving force later on. This aspect has been explored in section 3.4.

\subsection{Effect of pulsing $h_{ac}$ on $\chi^{\prime}(T)$ and the notion of the coexistence phase}

The $\chi^{\prime}(T)$ curves presented Fig.~4 is an attempt to explore the issue stated above. The solid and dotted curves represent the $\chi^{\prime}(T)$ data in the ZFC and the FCW modes recorded in $h_{ac}$ of 0.65~Oe. In the $\chi^{\prime}(T)$ curves, several temperatures were pre-selected (as marked by arrows) to impose an $h_{ac}$ pulse of 3~Oe  for a short duration ($\sim 1 sec$). As can be noted in Fig.~4, the three pre-selected temperatures are: (a) $T_X \rightarrow T_{pl1}$, (b) $T_Y > T_{pl2}$, and (c) $T_Z \rightarrow T_{pl3}$. At each of these chosen temperatures, an $h_{ac}$ signal of 3~Oe  at 211~Hz is momentarily applied ($ < $1~s duration) and $\chi^{\prime}(T)$ response is then measured with an $h_{ac}$ of 0.65~Oe. Application of an external drive just prior to $T_{pl1}$ results in a sharp drop in the diamagnetic response, implying a large reduction in $J_c(H)$ and, consequently, a significant enhancement in the correlation volume $V_c$ of the state having been subjected to a driving force as compared to that of the initial vortex state on the ZFC warm-up path. This is an interesting observation, as it is usually believed \cite{14,25} that the vortex states obtained via the ZFC mode are close to the equilibrium configuration.  The notion that the state of order in the ZFC mode can be further improved by momentary exposure to an external driving force was not anticipated. The better ordered state starts to disorder as the temperature is enhanced further across  $T_{pl1}$. No sudden change in $\chi^{\prime}(T)$ response is observed at $T_{pl2}$, however, at $T_{pl3}$, a sharp increase in the diamagnetic response occurs. Further increase in the temperature produces traversal of $\chi^{\prime}(T)$ response along a path different from that obtained for the ZFC and the FCW modes. At $T > T_{pl4}$, the newer path merges into those sketched for the ZFC and the FCW modes in Fig.~4. When the sample is exposed to an external drive of 3~Oe  at $T_Y (> T_{pl2})$, the enhancement in the ordered state does not result in a configuration better than that obtained while warming up the state obtained after pulsing an $h_{ac}$ at $T_X$. The path followed on warming up by the state produced at $T_Y$ is very different from that produced at $T_X$ in the temperature interval $T_{pl3}$ to $T_{pl4}$. However, while approaching the limiting temperature $T^{\star}$ ($> T_{pl4}$), the different $\chi^{\prime}(T)$ curves appear to merge into each other. It can further be noted that if $h_{ac}$ of 3~Oe  is pulsed at $T = T_Z$ ($\rightarrow T_{pl3}$), one can witness the creation of a vortex state with diamagnetic response lower than those of the states described above. As this newer state is warmed up, the process of disordering commences concurrently and the measured $\chi^{\prime}(T)$ curve follows another path until one reaches the limiting temperature $T^{\star}$. A comparison of $\chi^{\prime}(T)$ data in Fig.~3 and Fig.~4 implies that the temperature $T^{\star}$ also identifies the limit above which $\chi^{\prime}(T)$ response is independent of $h_{ac}$ in the interval 0.65~Oe  to 2~Oe, in which the data are being presented here.

The myriad of criss-crossing $\chi^{\prime}(T)$ curves with multi-peak structures relating to the ZFC mode in Fig.~4 bears a striking resemblance to the set of curves obtained by Marchevsky {\it et al} \cite{18} through local scanning ac Hall bar microscopy study on a weakly pinned single crystal of $2H$-NbSe$_2$ at $H = 270$~Oe ($\| c$) (cf. figures 2 and 3 in Ref. \cite{18}). Their study had revealed the coexistence of weaker and stronger pinned regions on a mesoscopic scale across the temperature region of PE anomaly. They have collated the ac susceptibility curves in different local regions and compared them with the average response of a macroscopic part of the sample (cf. Fig.~2 of Ref. \cite{18}). It is surmised that the topology of the interface separating the stronger and the weaker pinned pockets, their sizes, etc. have a complex relationship with the history of the dc magnetic field, amplitude of the ac drive and the disorder injected through the irregular edges of a given sample (shape and geometry). Different local regions in a sample can have differing distribution of coexisting weaker and stronger pinned vortex phases at a given ($H, T, h_{ac}$) and their temperature evolutions across the PE region take different paths. However, the (onset) position and the (temperature) width of the PE region were found to be independent of the $h_{ac}$ drive.  In the context of the notion of phase coexistence, our data in Fig.~4 implies that the entire sample could get filled with differing distributions of coexisting regions depending upon the thermomagnetic history of the $h_{ac}$ drive and thereby generate criss-crossing $\chi^{\prime}(T)$ curves between the onset ($T_{pl1}$) and the end position ($T^{\star}$) of the coexistence region. The PE is considered to imbibe the superheating/supercooling characteristic of a first order transition between the weaker (ordered) and stronger (disordered) pinned states. Above the onset temperature of the PE, the superheated ordered states are expected to transform to stronger pinned disordered states before the penultimate collapse of the pinning, which in our present sample appears to precipitate at $T_{pl4}$.  The local and global $\chi^{\prime}(T)$ data of Marchevsky {\it et al} \cite{18} in $2H$-NbSe$_2$ also conveys the same trend (cf. their Fig.~2 and our Fig.~4). The $\chi^{\prime}(T)$ data in Fig.~1 of Ref. \cite{18} had also shown that the coexistence characteristic lasts up to a limiting temperature higher than that at which the pinning collapses.  If we accept the suggestion \cite{18} that the history dependence in $\chi^{\prime}(T)$ response is related to the notion of coexisting phases, we may assert that the limiting value $T^{\star}$ in Fig.~4 identifies the spinodal temperature above which the weakly pinned ordered state does not superheat any more. It is also fruitful to  recall now the inference drawn by Thakur {\it et al} \cite{19,20} via their explorations of the metastability effects and the spinodal temperature at higher fields ($>1$~kOe) in single crystals of $2H$-NbSe$_2$ which belong to the same genre as studied by Marchevsky {\it et al} \cite{18}. At such fields ($>1$~kOe), these crystals display the second magnetization peak (SMP) anomaly prior to the PE in the isofield scans and the phase coexistence region exists between the onset temperature of the SMP anomaly ($T_{SMP}^{on}$) and the limiting temperature $T^{\star}$. Thakur {\it et al} \cite{19,20} have surmised that the phase coexistence region can be subdivided such that in the part adjoining the $T_{SMP}^{on}$, the ordered regions dominate and in the later part adjoining $T^{\star}$, the disordered regions take over that role. We have found an analogous behavior in the Yb$_3$Rh$_4$Sn$_{13}$ crystal under investigation from the results of $\chi^{\prime}(T)$ measurements in which the temperature is retraced from different positions in the interval $T_{pl1}$ to $T_{pl4}$. These are dealt with in the next section.

\subsection{Thermal cycling from different temperatures across the coexistence regime}

Figure~5 summarizes the $\chi^{\prime}(T)$ data obtained by retracing the temperature from five different positions along the ZFC $\chi^{\prime}(T)$ warm-up cycle. The solid curve in Fig.~5 identifies the ZFC $\chi^{\prime}(T)$ warm-up plot, on which the temperatures $T_{pl1}$ to $T_{pl4}$ have been identified. The chosen five temperatures from which the cool-down is  initiated are: (i) $T_I < T_{pl1}$, (ii) $T_{pl1} < T_{II} < T_{pl2}$, (iii) $T_{pl2} < T_{III} < T_{pl3}$, (iv) $T_{pl3} < T_{IV} < T_{pl4}$, and (v) $T_{V} > T_{pl4}$. For the purpose of  reference, the thin broken curve and the thick broken curve in Fig.~5 identifies the FCW and FCC $\chi^{\prime}(T)$ plots, respectively  initiated from 4.2~K. The $\chi^{\prime}(T)$ behavior recorded during the cool-down cycles can be  noted from data points having different colors and symbols.
Note first that on cooling down from $T_{I}$, i.e., before the entry into the PE region, $\chi^{\prime}(T)$ retraces its path as the data points lie on the ZFC warm-up curve. As the chosen temperature (viz., $T_{II}$) crosses $T_{pl1}$, the $\chi^{\prime}(T)$ fails to retrace its path on cooling below $T_{pl1}$ (see solid triangle data points). However, on further cooling down, the $\chi^{\prime}(T)$ response merges into the ZFC warm-up curve near 5~K.
Somewhat similar situation prevails even when the cool-down is initiated from $T_{III}$; $\chi^{\prime}(T)$ response (open circles) shows some hysteresis between $T_{pl2}$ and $T_{pl1}$, however, on crossing $T_{pl1}$, the response abruptly drops to a value nearly identical to that in the previous case and thereafter, $\chi^{\prime}(T)$ data points overlap with those recorded while cooling down from $T_{II}$. Next, when the cool-down gets initiated from $T_{IV}$ ($> T_{pl3}$), the $\chi^{\prime}(T)$ retraces the path down to $T_{pl3}$, but, thereafter, it suddenly drops to a value close to that obtained in the FCC run (as in Fig.~2). On further cooling down it follows a path nearly parallel to FCC run, which is well below that obtained during the FCW mode. Finally, as cooling is initiated from $T_{V}$ ($> T_{pl4}$), one obtains a $\chi^{\prime}(T)$ response identical to the FCC response (thick broken curve).

On the basis of $\chi^{\prime}(T)$ responses during the thermal cyclings initiated from ZFC warm-up path, the entire temperature regime can be subdivided into three parts: (a) $T < T_{pl1}$, (b) $T_{pl1} < T < T_{pl3}$, and (c) $T > T_{pl3}$. In the first case, $\chi^{\prime}(T)$ response of the ordered state is retraceable, in the second case, the coexistence phase traverses towards the (equilibrium) ordered phase ($T < T_{pl1}$) in the presence of a driving effect of the $h_{ac}$, and in the last case, the coexistence phase transforms towards a metastable disordered phase ($T < T_{pl1}$). We, thus believe that in the coexistence phase, the ordered regions dominate between $T_{pl1}$ and  $T_{pl3}$ and the disordered pockets dictate the behavior above $T_{pl3}$.

The metastable disordered phase at $T < T_{pl1}$ in Yb$_3$Rh$_4$Sn$_{13}$ can be driven to the ordered state under the influence of a suitably large $h_{ac}$. This switching between the disordered and the ordered states was first pointed out from bulk ac susceptibility measurements in crystals of $2H$-NbSe$_2$ and CeRu$_2$ by Banerjee {\it et al} \cite{25} and later confirmed also by local Hall bar ac susceptibility response in $2H$-NbSe$_2$ by Marchevsky {\it et al} \cite{18}, which overcomes the possible complications due to disorder injected through the edges. In the latter study, it was noted that when $h_{ac}$ was reduced from 30~mOe to 5~mOe, the FCC state did not display the PE feature, whereas  the PE peak is observed in the ZFC warm-up mode over the entire range of $h_{ac}$. The difference between the ZFC and FCC  $\chi^{\prime}(T)$ responses below the coexistence region is reported \cite{18} to be negligible when $h_{ac}$ is 30~mOe, and we may state that it would be substantial for $h_{ac} \approx 5$~mOe.

\subsection{Noise measurements in the ac susceptibility response and the spinodal temperature}

Noise signal in the ac susceptibility measurements is believed \cite{14} to arise from the possibility of transformations in the metastable vortex configurations under the given conditions of the experiment. Thakur {\it et al} \cite{20} have recently shown that the noise signal can be conveniently employed to ascertain the spinodal temperature $T^{\star}$ of the order-disorder transition as the $J_c(H)$ becomes a single valued function of $H$ above this temperature. Figure~6 displays the data on the measurement of noise in $\chi^{\prime}(T)$ at an applied field of 8~kOe and $h_{ac}$ of 0.65~Oe in Yb$_3$Rh$_4$Sn$_{13}$. The noise signal has been recorded with a Stanford Research Inc. (Model SR 850) lock-in amplifier having a flat band filter option, while keeping the synchronous filter off. The fluctuations in $\chi^{\prime}(T)$ correspond to measuring the standard deviation of the $\chi^{\prime}(T)$ signal, when filtering time of the lock-in amplifier has been made very small. The noise signal in $\chi^{\prime}(T)$ starts rising on going across $T_{pl1}$, its further increase fingerprints the abrupt changes across $T_{pl2}$ and $T_{pl3}$. The noise reaches the peak level just before the arrival of $T_{pl4}$, and thereafter it rapidly goes down, reaching the background level at the limiting temperature $T^{\star}$ as identified in Fig.~4. The noise signal indeed reflects the underlying changes in transformations amongst the metastable states and the overall glassiness in the vortex state from $T_{pl1}$ to $T_{pl4}$.
If $n$ is the number of fluctuations, then the noise in $\chi^{\prime}(T)$ is expected to be proportional to $n(n-1)/n^3$, arising from the number of accessible metastable configurations at a given ($H, T, h_{ac}$). The decrease in noise across $T_{pl4}$ is probably caused by the phase cancellation of a large number of incoherent fluctuations in the nearly amorphous phase of the vortex solid just above this temperature. Once the $J_c(H)$ becomes path independent ($n \approx 1$) at the spinodal temperature $T^{\star}$, the noise signal would be expected to recede to the background level.

\section{Summary}

We have presented results of investigations on the metastability effects related to the PE phenomenon in a weakly pinned crystal of Yb$_3$Rh$_4$Sn$_{13}$. The PE anomaly in the given sample surfaces in a conspicuous manner in a field of 5~kOe (where inter-vortex spacing $a_0 \sim 700$~\AA), whereas the range of interaction between the vortices (i.e., the penetration depth, $\lambda$) in this compound is estimated to be about 2200~\AA.

The isofield scans in the ac susceptibility measurements in 5~kOe reveal the disordering of the ordered flux line lattice in a multi-step manner. The number of these steps enhances as the dc field progressively enhances. In a field of 10~kOe (where $a_0 \sim 500$~\AA), one can mark out four temperatures across which the partial fracturing of the ordered vortex domains occur. Earlier experiments \cite{15,16} in a weakly pinned crystal of isostructural Ca$_3$Rh$_4$Sn$_{13}$ had made us believe that retracing the temperature across a fracturing temperature would lead to progression towards a metastable disordered state. However, the results in the present Yb$_3$Rh$_4$Sn$_{13}$ crystal show that reversing the path across the first two step temperatures (viz., $T_{pl1}$ and $T_{pl2}$) elucidates the irreversible characteristic in the $\chi^{\prime}(T)$ response but it eventually progresses towards the ordered vortex state (as in the ZFC warm-up mode). Recalling the notion that the dislocations could get spontaneously injected \cite{26,27} into the ordered state at each fracturing step \cite{14}, the above stated behavior implies that some of these dislocations can be squeezed out \cite{17,27} on retracing the path while remaining under the influence of a driving force. The thermal cyclings across the step temperatures $T_{pl3}$ and $T_{pl4}$ do result in the progression of the (partially) fractured state towards the metastable disordered state (as in the FCW mode).

The disorder (pinning) induced order-disorder transition commencing at the onset temperature of the PE imbibes the supercooling/superheating notion usually considered a characteristic of the thermodynamic first order transition \cite{28}. Prior to the onset temperature of the PE anomaly, the metastable state can be (easily) realized as distinct from the (stable) ordered state. The phase coexistence of the ordered and the disordered regions happens from the onset position up to the spinodal temperature \cite{19,20}, above which the $J_c(H)$ is a history independent single valued function of $H$. The initial portion of the phase coexistence regime is such that the superheated ordered regions could dominate the overall response of the sample. Once the disordered regions start to percolate over the entire sample, attempts to retrace the path in a thermal cycling result in transformation of the superheated ordered pockets towards the disordered state, and the sample remains filled with metastable disordered states thereafter on cooling down (below T$_{pl1}$). Such a metastable state (at $T~<~T_{pl1}$) can be driven into the ordered state prior to the onset temperature of the PE by imposition of a suitably large driving force \cite{25}.

We had earlier (cf. Fig.~3 of Ref. \cite{29}) drawn attention to the similarities in the vortex phase diagrams in weakly pinned crystals of Yb$_3$Rh$_4$Sn$_{13}$ and Ca$_3$Rh$_4$Sn$_{13}$. The ($T_{pl}$, $H_{pl}$) values in these diagrams denote the first order transition like order-disorder transition line. At the low field end, this line appears to terminate in a critical point for which the field value lies between 3 to 4~kOe interval, where $\lambda / a_0$ for the above two compounds is greater than two and one expects interaction effects between the vortices to be significant. It may be useful to recall at this juncture that Park {\it et al} \cite{30} carried out small angle neutron scattering experiments in a crystal of Nb, while concurrently recording the ac susceptibility data in it. They find that the Bragg glass phase persists below the field values (viz., 0.8~kOe), where they cannot observe PE in their sample. In their sample, the surface superconductivity is also present between $H_{c2}(T)$ and $H_{c3}(T)$, and the PE line ($T_p(H)$), $H_{c2}(T)$ and $H_{c3}(T)$ lines are argued to meet at a multiple critical point in the Bragg glass phase. We would like to argue that the first order like $T_{pl}(H)$ line in Yb$_3$Rh$_4$Sn$_{13}$ crystal terminates at a critical point in the elastically deformed Bragg glass phase. The lower field limit for the Bragg glass phase in the ternary stannide crystals may be in the range of $\lambda / a_0 \sim 1$ (i.e., about 0.5~kOe). The single vortex pinning limit in Ca$_3$Rh$_4$Sn$_{13}$ was estimated to be $\sim 100$~Oe. It is likely that between 0.1 to 0.5~kOe, the multi-domain (poly-crystalline) vortex glass regime prevails. The issue of critical points in the vortex phase diagrams of weakly pinned crystals of ternary stannides shall be dealt with in a comprehensive manner elsewhere \cite{31}.

\section{Acknowledgment}

Two of us (SS and ADT) would like to acknowledge the TIFR endowment fund for the Kanwal Rekhi career development support.

$^\ast$ Email: tomy@phy.iitb.ac.in~~:~~$^{\ast\ast}$ Email: grover@tifr.res.in

\newpage

\begin{figure} 
\includegraphics[scale=0.6,angle=0]{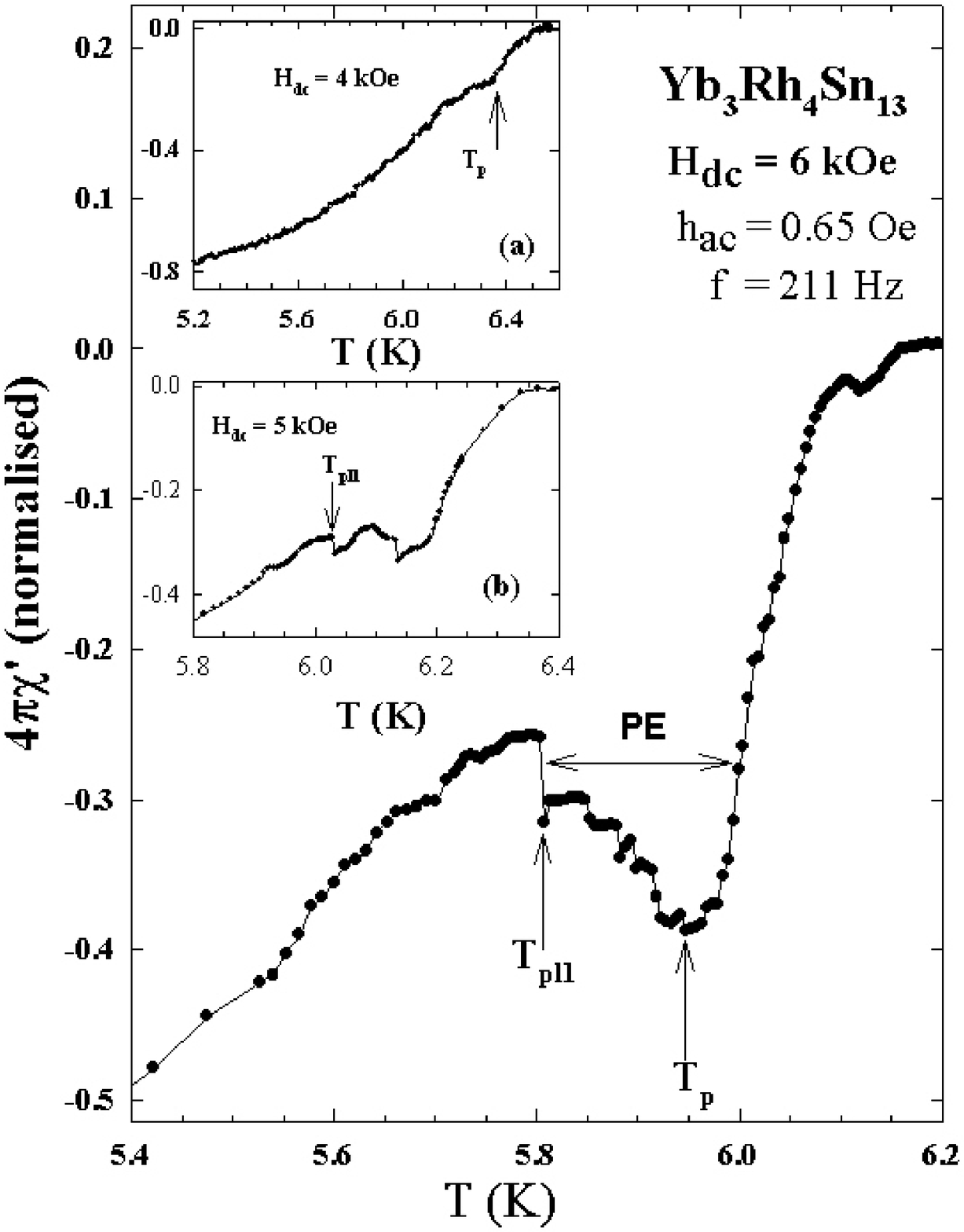}
\caption{Temperature dependence of the in-phase ac susceptibility $\chi^{\prime}(T)$ in a single crystal of (cubic) Yb$_3$Rh$_4$Sn$_{13}$ at the fields indicated. The inset panel (a) shows the PE anomaly at $T_p$ in a nascent form at $H_{dc} =$ 4~kOe. At 5~kOe, the PE anomaly can be recognized well (see panel (b)), it has a characteristic two peak structure which commences at $T_{pl1}$. The main panel shows that at 6 kOe, the structure across the PE region (between $T_{pl1}$ and $T_p$) displays several modulations.}
\end{figure}
\begin{figure} 
\includegraphics[scale=0.6,angle=0]{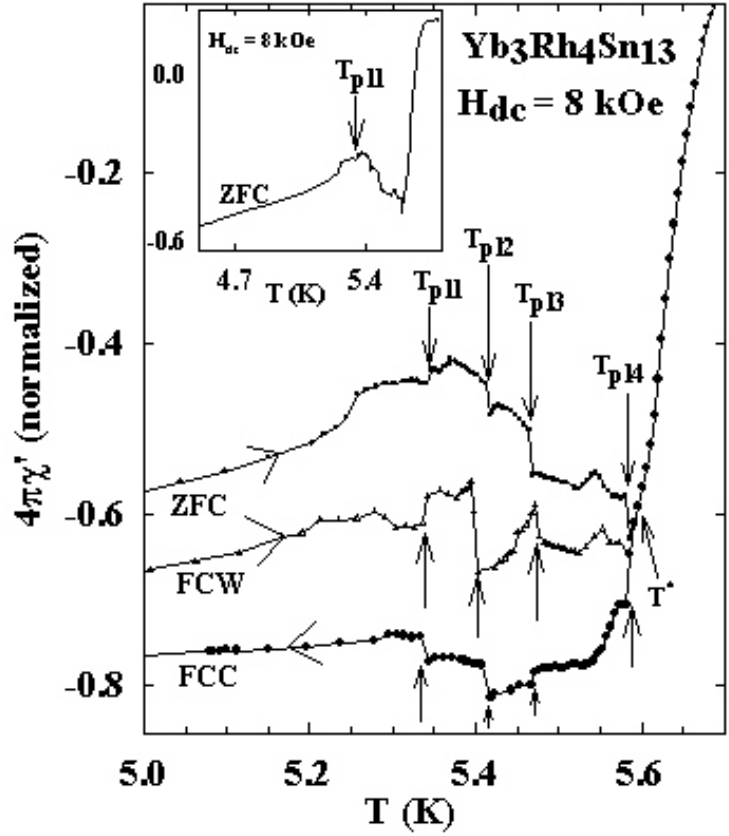}
\caption{The $\chi^{\prime}(T)$ plots (on an expanded scale) in a field of 8 kOe in the three modes, ZFC warm-up, FC cool-down and FC warm-up, in a crystal of Yb$_3$Rh$_4$Sn$_{13}$. The inset panel shows the PE anomaly in the full $\chi^{\prime}(T)$ curve for $h_{ac} = 0.65$~Oe in the ZFC mode. The four temperatures ($T_{pl1}$ to $T_{pl4}$) at which sharp changes (notion of fracturing) in $\chi^{\prime}(T)$ response occur have been marked for each of the curves. The $T^{\star}$ value has also been identified (see figures~3 and 4).}
\end{figure}
\begin{figure} 
\includegraphics[scale=0.6,angle=0]{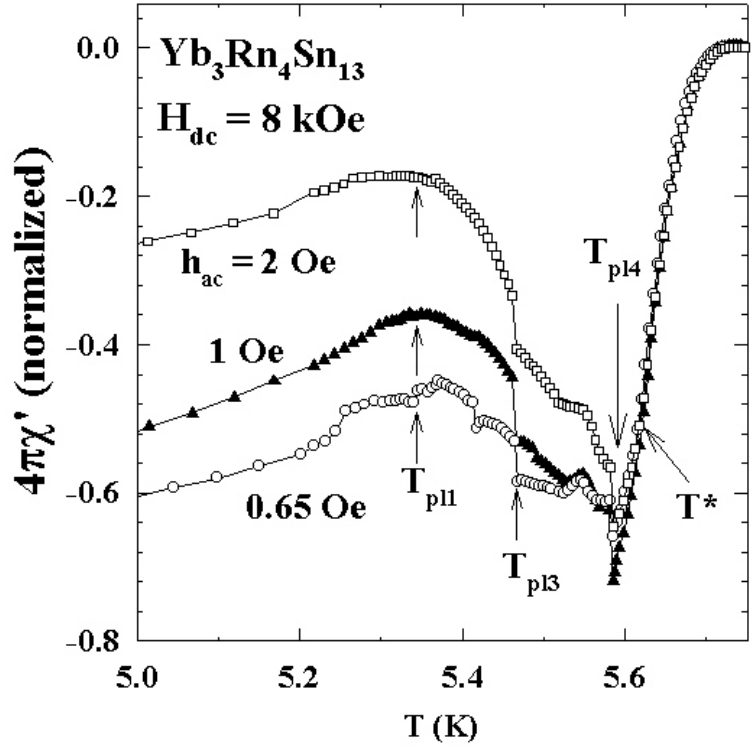}
\caption{Comparison of $\chi^{\prime}(T)$ curves obtained for different $h_{ac}$ values in a field of 8~kOe in a crystal of Yb$_3$Rh$_4$Sn$_{13}$. The fracturing temperatures $T_{pl1}$, $T_{pl3}$ and $T_{pl4}$ have been marked. Note that all the three curves appear to merge above the limiting temperature $T^{\star}$.}
\end{figure}
\begin{figure} 
\includegraphics[scale=0.6,angle=0]{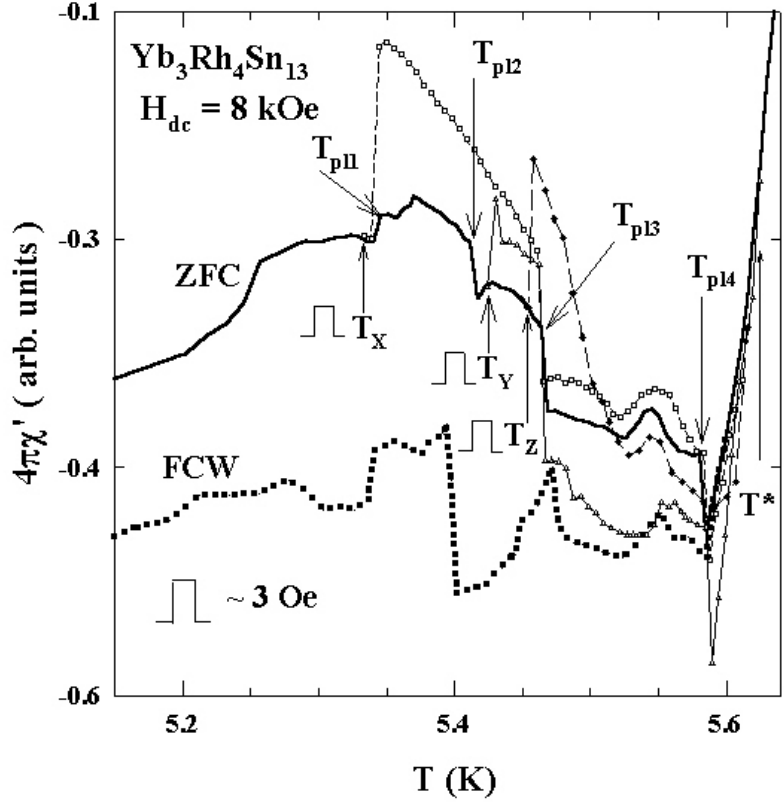}
\caption{The $\chi^{\prime}(T)$ data in $h_{ac}$ of 0.65 Oe in a field of 8~kOe in Yb$_3$Rh$_4$Sn$_{13}$, showing the effect of exposing the sample to a larger $h_{ac}$ of 3~Oe at three pre-selected temperatures as indicated: (i) $T_X \rightarrow T_{pl1}$, (ii) $T_Y > T_{pl2}$, and (iii) $T_Z \rightarrow T_{pl3}$. After an exposure to larger $h_{ac}$, the $\chi^{\prime}(T)$ data are obtained in $h_{ac}$ of 0.65 Oe during further warm-up. The solid and the dotted curves in this figure identify the ZFC warm-up and the FC warm-up plots of Fig. 2 for the purpose of reference. The limiting temperature $T^{\star}$, above which the path dependence in $\chi^{\prime}(T)$ behavior ceases has been identified.}
\end{figure}
\begin{figure} 
\includegraphics[scale=0.6,angle=0]{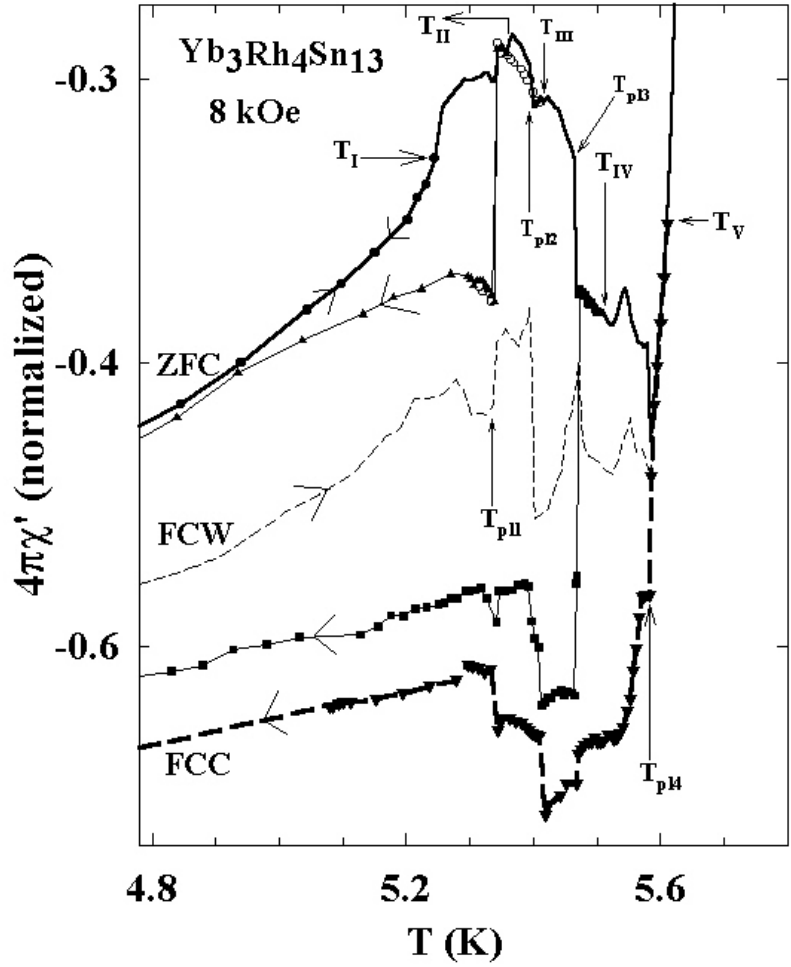}
\caption{The $\chi^{\prime}(T)$ data in a field of 8 kOe in Yb$_3$Rh$_4$Sn$_{13}$ showing the effect of thermal cyclings initiated from five selected temperatures as indicated: (i) $T_I < T_{pl1}$, (ii) $T_{pl1} < T_{II} < T_{pl2}$, (iii) $T_{pl2} < T_{III} < T_{pl3}$, (iv) $T_{pl3} < T_{IV} < T_{pl4}$, and (v) $T_{V} > T_{pl4}$.}
\end{figure}
\begin{figure} 
\includegraphics[scale=0.6,angle=0]{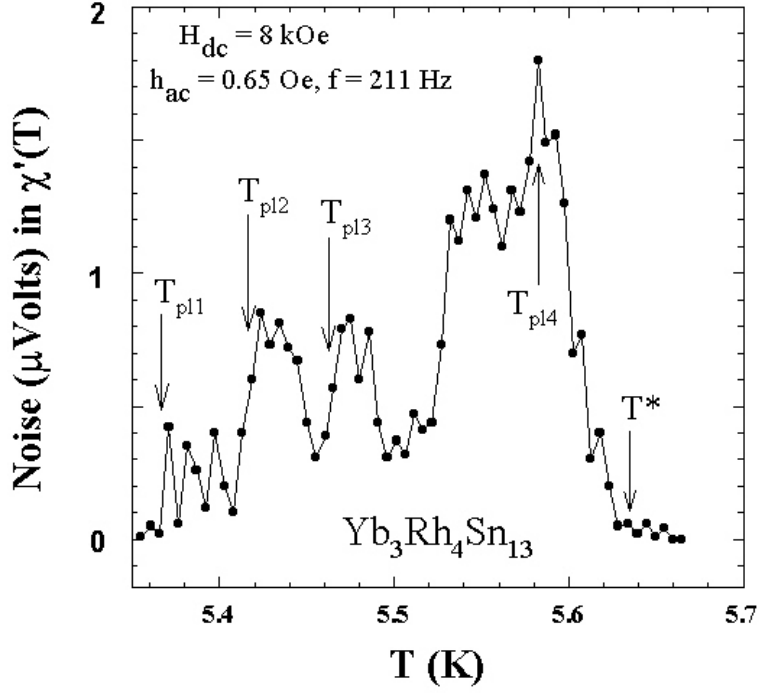}
\caption{The noise data in $\chi^{\prime}(T)$ recorded in $h_{ac}$ of 0.65 Oe at $f~=~211$~Hz at $H_{dc}~=~8$~kOe in the crystal of Yb$_3$Rh$_4$Sn$_{13}$ (see text for details). The fracturing temperatures ($T_{pl1}$ to $T_{pl4}$) and the $T^{\star}$ value (cf. figures 3 and 4) have been marked.}
\end{figure}
\end{document}